\documentclass[letterpaper,prl,twocolumn]{revtex4}
\usepackage{graphicx}
\usepackage{dcolumn}
\usepackage{bm}
\usepackage{color}
\usepackage{soul}
\usepackage[english]{babel}
\usepackage{epsfig}
\usepackage{amsmath}
\usepackage{amsfonts}
\usepackage{amssymb}
\usepackage{float}
\usepackage{mathrsfs}

\begin{document}

\title{Bound state soliton gas dynamics underlying the noise-induced modulational instability}

\author{Andrey Gelash$^{1,2}$}
\author{Dmitry Agafontsev$^{1,3}$}
\author{Vladimir Zakharov$^{1,4}$}
\author{Gennady El$^{5}$}
\author{St\'ephane Randoux$^{6,7}$}
\author{Pierre Suret$^{6,7,*}$}

\affiliation{$^{1}$Skolkovo Institute of Science and Technology, Moscow, 143026, Russia}
\affiliation{$^{2}$Novosibirsk State University, Novosibirsk, 630090, Russia}
\affiliation{$^{3}$P.\,P. Shirshov Institute of Oceanology, RAS, 117218, Moscow, Russia}
\affiliation{$^{4}$P.\,N. Lebedev Physical Institute, 53 Leninsky ave., 119991, Moscow, Russia}
\affiliation{$^{5}$Northumbria University, Newcastle upon Tyne, United Kingdom}
\affiliation{$^{6}$Laboratoire de Physique des Lasers, Atomes et Molecules, UMR-CNRS 8523,  Universit\'e de Lille, France}
\affiliation{$^{7}$Centre d'Etudes et de Recherches Lasers et Applications (CERLA), 59655 Villeneuve d'Ascq, France}

\email[Corresponding author : ]{Pierre.Suret@univ-lille.fr}

\begin{abstract}
  {We investigate theoretically the fundamental phenomenon of the spontaneous, noise-induced modulational instability (MI) of a plane wave. The long-term statistical properties of the noise-induced MI have been previously observed in experiments and in simulations but have not been explained so far. In the framework of inverse scattering transform (IST), we propose a model of the asymptotic stage of the noise-induced MI based on  $N$-soliton solutions ($N$-SS) of  the integrable focusing one-dimensional nonlinear Schr\"odinger equation (1D-NLSE).  These $N$-SS are  bound states of strongly interacting solitons having a specific distribution of the IST eigenvalues together with random phases. We use a special approach to construct ensembles of multi-soliton solutions with statistically large number of solitons $N\sim100$. Our investigation demonstrates complete agreement in spectral (Fourier) and statistical properties between the long-term evolution of the condensate perturbed by noise  and the constructed multi-soliton bound states. Our results can be generalised to a broad class of integrable turbulence problems in the cases when the wave field dynamics is strongly nonlinear and driven by solitons.}
\end{abstract}

\maketitle

Integrable partial differential equations (PDE) such as Sine-Gordon,  Korteweg-de Vries (KdV)  and the one-dimensional nonlinear Schr\"odinger equation (1D-NLSE) are considered as universal models of nonlinear physics ~\cite{Yang}. They describe at leading order various nonlinear systems and can be integrated using the Inverse Scattering Transform (IST), often seen as a nonlinear analogue of the Fourier transform~\cite{Zakharov:72,Zakharov:74, Ablowitz:74,Zakharov:79, Ablowitz:81}. In the IST theory, the  wave field plays the role of a potential in a linear scattering problem associated with the nonlinear PDE. The traditional IST theory deals with decaying potentials ~\cite{Ablowitz:81,Novikov},  but constant non-zero boundary conditions  at infinity can also be considered ~\cite{Demontis:13}.  An extension of the IST method to the case of periodic boundary conditions  is also available in the framework  of the so-called {\it finite gap theory} (FGT) \cite{Belokolos, Tracy:88}. 

In contrast to  deterministic initial conditions, the propagation of {\it random waves} in integrable systems  is an open theoretical problem of significant applied interest due to complexity of many real world nonlinear phenomena modeled by integrable equations. For this reason, {\it Integrable Turbulence} (IT) has been recently introduced as a  ``new chapter of turbulence theory'' by V. Zakharov~\cite{Zakharov:09} and is now an active theoretical and experimental field of research ~\cite{Agafontsev:15, Onorato:16,  SotoCrespo:16, Randoux:16, Roberti:19}. The central question in integrable turbulence is the evolution of the statistical properties of a random  wave field  in the course  of its propagation through a nonlinear dispersive medium.  In particular, optics and hydrodynamics provide very favorable settings for the investigation of integrable turbulence because the nonlinear propagation of one-dimensional waves in water tank or in an optical fiber are described at leading order by the 1D-NLSE or KdV models~\cite{Walczak:15,Suret:16,Tikan:18,Koussaifi:18,Cazaubiel:18}.

It has been realised in \cite{Zakharov:MI:09} that the development of the noise-induced  {\it Modulational Instability} (MI, also known as the Benjamin-Feir instability) arising in the focusing regime of the 1D-NLSE represents a prominent example of IT phenomena.
This instability can be observed in many physical systems such as deep water waves~\cite{Osborne}, BEC~\cite{Strecker:02} or nonlinear optical waves~\cite{Agrawal}. In the traditional formulation, the development of MI is  seen as the amplification of an initially small sinusoidal perturbation of a plane wave  -- the {\it condensate} \cite{Zakharov:MI:09,Tracy:88}. In this case, the nonlinear stage of MI is  described by exact solutions of the 1D-NLSE -- Akhmediev Breathers~\cite{Akhmediev:85,Akhmediev:86,Akhmediev:09, Grinevich:18}.

When the small initial perturbation of the condensate is a  random process, the  numerical simulations  of the focusing 1D-NLSE show that the long-time evolution is characterised by a stationary  single-point statistics which is  Gaussian despite the presence of random highly nonlinear breather structures ~\cite{Agafontsev:15,Akhmediev:IntegrableTurbulence:16,SotoCrespo:16}. It has also been shown recently that this long-time (stationary) statistics is characterized by a quasi-periodic structure of the spatial autocorrelation function of the wave field intensity~\cite{Kraych:19bis}. 

While all these remarkable features of IT have been recently demonstrated experimentally by using an optical fiber loop~\cite{Kraych:19bis}, they are still not understood theoretically.

Integrable turbulence can be approached from a completely different perspective which is close to classical statistical mechanics. In 1971, V. Zakharov introduced the concept of soliton gas as an infinite collection of interacting KdV solitons that are randomly distributed in space \cite{Zakharov:71:kinetic}.  Originally introduced for the case of small density due to significant simplifications in the analytical treatment, the notion of soliton gas has been  extended to  gases of finite density both for KdV and for the focusing 1D-NLSE  \cite{El:05:prl}.  Importantly, the macroscopic properties of dense soliton gases are determined by pairwise collisions of solitons accompanied by phase shifts that are accumulated at long time leading to significant corrections to the average soliton velocities.
The key role in the soliton gas theory is played by the spectral (IST) distribution function which has the meaning of the density of states $f(\lambda, x, t)$, so that $f(\lambda_0, x_0 ,t)d\lambda dx$ is the number of solitons with the spectral parameter $\lambda$ in the interval $[\lambda_0, \lambda_0+ d\lambda]$ found in the space interval $[x_0, x_0+dx]$. For  spatially non-uniform soliton gas the evolution of $f$ is described by a kinetic equation \cite{Zakharov:71:kinetic, El:05:prl}.

Given the above two approaches to IT one can naturally pose a question about a possibility of  describing the development of modulational instability of a plane wave by considering the  soliton gas dynamics for some special  spectral (IST) distribution. In fact, the  idea to explain nonlinear stage of MI using soliton interactions has been put forward as early as in 1972   by  V. Zakharov and  A. Shabat ~\cite{Zakharov:72}. However, rather paradoxically, up to now, this possible link between soliton interactions and MI of a plane wave has not been explored. 

In this paper, we  provide a bridge between the  two fundamental phenomena of nonlinear physics  by showing that soliton gas dynamics explains the fundamental features of the nonlinear stage of the noise-induced modulational instability of a plane wave.   More precisely, we demonstrate a remarkable agreement between the  spectral (Fourier) and statistical properties of an unstable plane wave in the long-time evolution and those of a soliton gas representing a random infinite-soliton bound state (i.e. the 1D-NLSE solution in which all solitons  are stationary in an appropriate reference frame). The two key ingredients in our analysis are (i) a special choice of the spectral (IST) distribution in the soliton gas and  (ii) random phases of the so-called norming constants.

\medskip

We consider the focusing 1D-NLSE in the standard dimensionless form: 
\begin{equation}\label{eq:NLSE}
  i\frac{\partial \psi}{\partial t} + \frac{1}{2}\frac{\partial^2 \psi}{\partial x^2} + |\psi|^2\psi = 0\,,
\end{equation}
where $\psi(t,x)$ represents the complex field, and $x$ and $t$ are space and time.  The plane wave solution of Eq.~(\ref{eq:NLSE}) is $\psi_c(t,x) = A\,\exp{i A^2t}$, where $A$ is the condensate amplitude. Without loss of generality, here we assume that $A=1$ . The classical formulation of the noise-induced MI problem is to consider the initial condition composed of the condensate with some additional noise:
\begin{equation}\label{eq:IC}
 \psi(t=0,x) = 1 + \eta(x)\,,
\end{equation}
where $\eta$ is the noise with $\langle \eta \rangle=0$  and $\sqrt{\langle |\eta|^2 \rangle} \ll 1$. The condensate is unstable with respect to long-wave perturbations with the growth rate $\Gamma(k)=k\sqrt{1-k^2/4}$. In the following, the spectral width of the the noise is assumed to be much larger than the wave number $k_m=\sqrt{2}$ corresponding to the maximum gain of the modulational instability. This problem has been widely investigated in recent works by using  periodic boundary conditions in a box of large size~\cite{Dudley:14, Toenger:15, Agafontsev:15,Kraych:19}.\\

Nonlinear wave fields ruled by Eq.~(\ref{eq:NLSE}) can be characterized by the so-called  scattering data (which we shall call the {\it IST spectrum}). In the general case,  the IST spectrum of spatially localized wave fields $\psi(x)$ -- with zero boundary conditions -- consists of the {\it continuous and discrete} components.  A special class of solutions, the {\it N-soliton solutions} ($N$-SS),  exhibits only discrete spectrum consisting of $N$ discrete complex-valued eigenvalues $\lambda_{n}$, $n=1,...,N$, and complex coefficients $C_{n}$ (norming constants) defined for each $\lambda_{n}$. The key result of IST theory is that, while the wave field $\psi$ exhibits a complex dynamics, the IST spectrum changes trivially in time ~\cite{Novikov}: 
\begin{equation}\label{ScattData(t)}
\forall n: \lambda_{n}=\mathrm{const},\quad C_n (t) = C_n (0) e^{-2i \lambda_n^2 t}\,.
\end{equation}

The asymptotic evolution of the N-SS for $t \to \infty$ generally leads to a superposition of $N$ moving solitons (1-soliton solutions). Each soliton corresponds to a point $\lambda_i$ of the discrete spectrum so that  $\mathrm{Im} \lambda_i$ is proportional to the soliton's amplitude and $\mathrm{Re} \lambda_i$ -- to its velocity. 
The value $C_i$ determines the soliton phase and its position.

There is a special class of N-SS's with all $\mathrm{Re} \lambda_i =0$, the so-called bound states. In the following, we consider an ensemble of N-SS's  with the  eigenvalues located on the imaginary axis and random phases in Eq.~\eqref{ScattData(t)}, i.e. $C_n(0)=|C_n(0)| e^{i\theta_n}$, where $\theta_n$ are uniformly distributed in $[0, 2\pi)$.   For $N \gg 1$, we assume the following (Weyl's) distribution of IST eigenvalues $\lambda_n = i\, \beta_n$:

\begin{equation}\label{eq:EIGdistr}
f(\beta) = \beta/(\sqrt{1-\beta^2}) 
\end{equation}
We shall call the limit at $N\to \infty$ of the described  random $N$-SS ensemble a {\it bound state soliton gas}. \\

 The fundamental conjecture  proposed and studied in this paper is that {\it  spectral (Fourier) and statistical properties of  the stationary state  of the noise-induced MI can be described by  a bound state soliton gas  with   certain  statistical distribution of the IST spectrum  consisting of the Weyl's distribution \eqref{eq:EIGdistr} of discrete eigenvalues  and random,  uniform distribution of the phases $\theta_n$}.  The assumption of random phase in the long-term evolution of a stochastic field is natural because  the phase rotations $-2i \lambda_n^2 t$ for large $t$  introduce an effective randomization of the phases.  Random phases of norming constants are proposed here to describe IT in the framework of IST;  similarly, the so-called ``random phase approximation'' in wave turbulence theory corresponds to random phases of the Fourier components of weakly dispersive waves~\cite{Zakharov, Nazarenko}. 

The motivation behind the statistical eigenvalue distribution \eqref{eq:EIGdistr} in the 1D-NLSE soliton gas is the Bohr-Sommerfeld quantization rule for the discrete spectrum of  the semi-classical Zakharov-Shabat scattering problem with the potential in the form of a real-valued rectangular box of unit amplitude and width $L_0 \gg 1$ (see~\cite{Zakharov:72,Novikov} and also~\cite{Lewis:85}):
\begin{equation}\label{eq:EIGfixed}
\lambda_n = i\, \beta_n = i\, \sqrt{1-\left[\frac{\pi(n-1)}{ L_0} \right]^2}, \quad n=1, 2, \dots, N,
\end{equation}
where $N=\hbox{int}[L_0/\pi]$. Importantly, for a "semi-classical'' box,  the contribution of the "non-soliton'' part of the  field, {\it i.e.} of the continuous IST spectrum, to the solution decays exponentially with $L_0$ and so can be neglected~\cite{Novikov}. 
The continuum limit of Eq. \eqref{eq:EIGfixed} with $N \to \infty$, $\beta_n \to \beta$ yields the Weyl's distribution \eqref{eq:EIGdistr} for $f(\beta)=\frac{1}{L_0}dn/d\beta$.\\

The derivation of the general $N$-SS of the 1D-NLSE is a classical result of the IST theory~\cite{Zakharov:72}. However, the numerical computation of $N$-solitons solutions of the focusing 1D-NLSE with large $N\sim 100$ has been realized for the first time in 2018~\cite{Gelash:18b}. Here, we use  the  approach developed in ~\cite{Gelash:18b} that is based on the specific implementation of the so-called dressing method~\cite{Zakharov:78} combined with 100-digits arithmetics (see ~\cite{Gelash:18b} for details). In the following, we shall compare the nonlinear stage of MI with the dynamical and statistical properties of the bound state soliton gas built as a random ensemble of $N$-solitons solutions with $N=128$ and the spectral distribution~\eqref{eq:EIGdistr}. In the results reported in this paper, the discretisation of the Weyl distribution \eqref{eq:EIGdistr} is taken at the Bohr-Sommerfeld quantization points \eqref{eq:EIGfixed}. Similar results are obtained when soliton eigenvalues are randomly distributed using the probability function~\eqref{eq:EIGdistr}.

We study $10^3$ realisations of $128$-SS with random uniform distribution of soliton phase parameters $\theta_n$ in the interval $[0,2\pi)$.
The density of the gas, {\it i.e.} the number of soliton per unit length, plays a crucial role in the dynamics. Our numerical investigations have shown that the higher  the density is, the better is the agreement between soliton gas and the stationary state of MI. In the dressing method, the density is empirically controlled by  ``space position parameters'' $x_{0n}$ and has its maximum (critical) density when all $x_{0n}=0$ and  all $|C_n|=1$. This critical density, which we observe empirically, coincides with the density of solitons corresponding to the Bohr-Sommerfeld quantization rule~\eqref{eq:EIGfixed}. Note that, when all $x_{0n}=0$, the $N$-SS solution is symmetric; in order to avoid this artificial symmetry, we use a random uniform distribution of space position parameters $x_{0n}$ in a narrow interval $[-2,2]$, so that the soliton density remains practically unchanged.

Finally, it is important to note that the norming constants used in the dressing methods somewhat differ from those appearing in the standard IST formulation (see e.g.\cite{Sun:16}). However, the uniform random phase distribution employed in our dressing construction of N-SS's also translates to uniform distribution for the traditional IST phases

\begin{figure}[h!]
\centering
\includegraphics[width=3.4in]{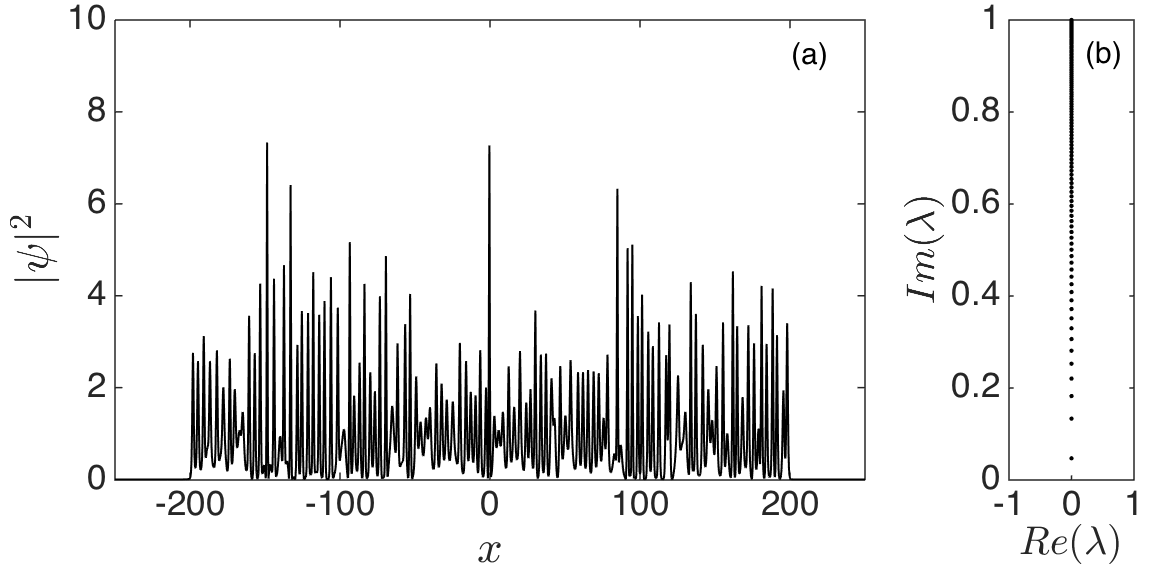}
\caption{\label{Figure01}
{\bf Example of one realization of 128-SS with random soliton phases.}
(a) Intensity profile $|\psi(x,t=0)|^2$.
(b) Soliton eigenvalues are computed from Eq.~(\ref{eq:EIGfixed}).
}
\end{figure}

\begin{figure}
\centering
\includegraphics[width=3.5in]{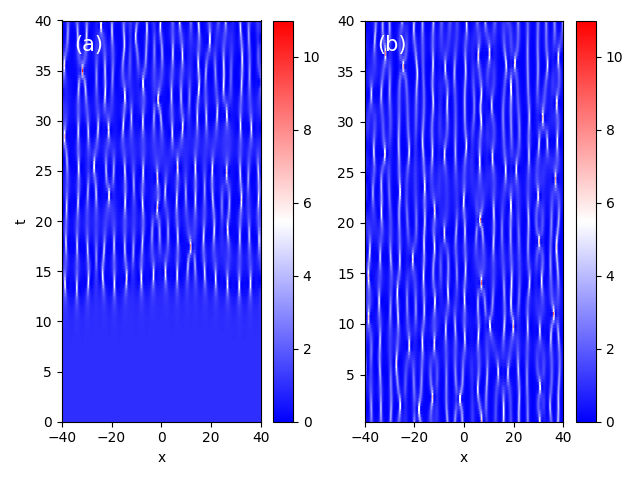}
 \caption{\label{Figure02}
{\bf Numerical simulations of 1D-NLSE : Space-Time diagrams of $|\psi(x,t)|^2$.} (a) Noise-induced Modulational Instability of a plane wave. (b) Dynamics of the random phase $N$-SS (the initial condition $|\psi(x,0)|^2$ is shown in the Fig.~\ref{Figure01}.a) }
\end{figure}

An example of the bound state N-SS with $N=128$ is displayed in Fig.~\ref{Figure01}. We first compare qualitatively the temporal evolution of this bound state soliton gas and the temporal evolution of an unstable plane waves (Fig.~\ref{Figure02}). We simulate the MI development using pseudospectral Runge-Kutta 4th-order method as described in~\cite{Agafontsev:15}. Periodic boundary conditions in a box of size $L\simeq 570$ are used and the initial conditions are given by Eq. \ref{eq:IC} with  $\langle|\eta|^2\rangle=10^{-5}$. In the spatio-temporal dynamics of the MI, one recognizes the emergence of the well-known structures resembling the Akhmediev-Breathers (Fig.~\ref{Figure02}.a). The spatio-temporal evolution of the bound state N-SS is also computed by using numerical simulations of 1D-NLSE (Fig.~\ref{Figure02}.b). Remarkably, the features characterizing the dynamics of the $N$-soliton and of the noise-induced MI  are qualitatively very similar. Note that, having purely imaginary eigenvalues, the N-SS used here are bound states and the solitons do not separate at long time.

We now compare quantitatively the statistical properties of soliton gas and of the nonlinear stage of MI. In particular, the long-term evolution of the noise-induced MI is characterised by stationary values of the potential and kinetic energy, kurtosis and also by stationary shapes of the (Fourier) spectrum, the probability density function (PDF) of the intensity $I=|\psi|^2$, and the autocorrelation function $g^{(2)}$ (see \cite{Agafontsev:15,Kraych:19}).

\begin{figure}
\centering
\includegraphics[width=0.99\linewidth]{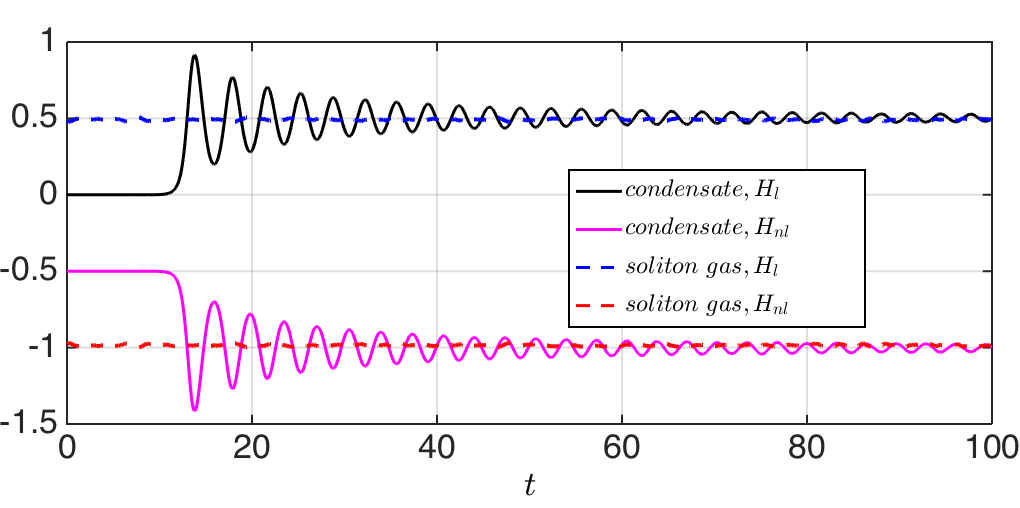}
 \caption{\label{Figure03}
The evolution of ensemble averaged kinetic $\langle H_{l}(t) \rangle$ and potential $\langle H_{nl}(t) \rangle$ energies for 
 the noise induced development of MI (black curves) and random phase $128$-SS (red curves).}
\end{figure}

The total energy (Hamiltonian) $E$ of the wave field is one of the infinite constants of motion of 1D-NLSE ~\cite{Novikov}:
\begin{eqnarray}\label{eq:Energy}
&& E = H_{l} + H_{nl},\quad H_{l}=\frac{1}{2} \frac{1}{L}\int_{-L/2}^{L/2}|\psi_{x}|^{2}\,dx,\nonumber\\
&& H_{nl} = - \frac{1}{2}\frac{1}{L}\int_{-L/2}^{L/2}|\psi|^{4}\,dx.
\end{eqnarray}
 In the case of MI, it has been shown that after some oscillatory transient, the kinetic energy reaches a stationary value $H_l=0.5$ while the potential energy reaches $H_{nl}=-1$ (see ~\cite{Agafontsev:15} and Fig.~\ref{Figure03}). Remarkably, the chosen bound state soliton gas is characterised by the same stationary values of $\langle H_l\rangle$ and $\langle H_{nl}\rangle$ (dashed lines in Fig.~\ref{Figure03}). Here the averaging $\langle \dots \rangle$ has been performed over $10^3$ random phase realisations. 

In the following, we perform ensemble averaging, together with temporal averaging, both for the noise-induced MI and for the $N$-SS. In the case of the condensate, the temporal averaging is performed when the system is sufficiently close (by its statistical properties) to the asymptotic stationary state ($t\in [160, 200]$). In the case of the $N$-SS, time averaging is started from the very beginning of the system evolution. It is extremely important to note that the time-averaging is used here only for practical reasons. Simulations made by averaging solely with  ensembles of realizations of random phases of the norming constants provide the same results.

\begin{figure}
\centering
\includegraphics[width=0.99\linewidth]{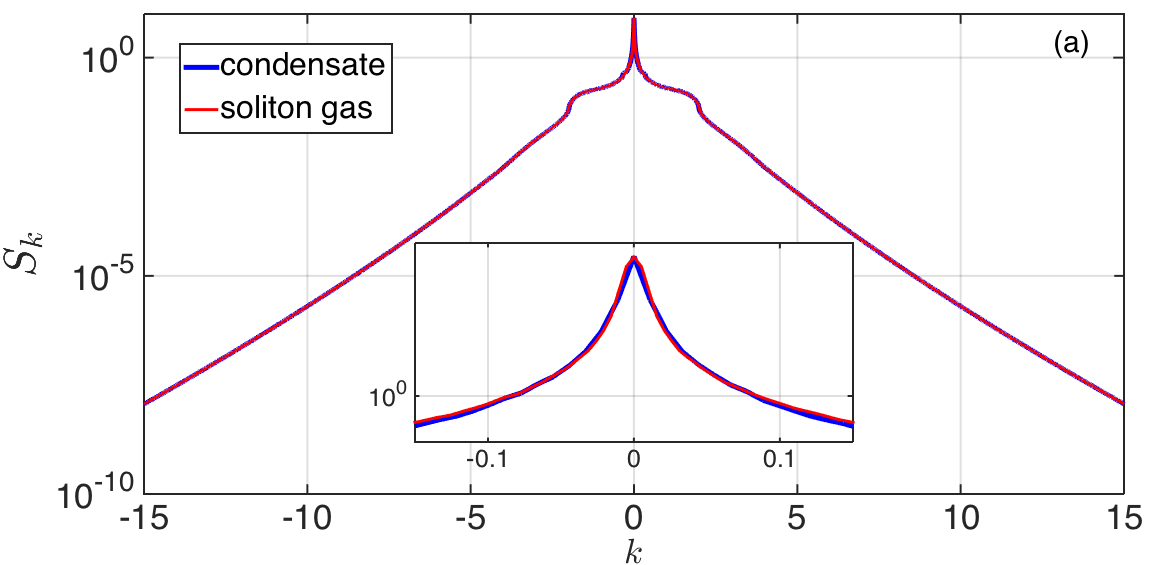}
\\
\includegraphics[width=0.99\linewidth]{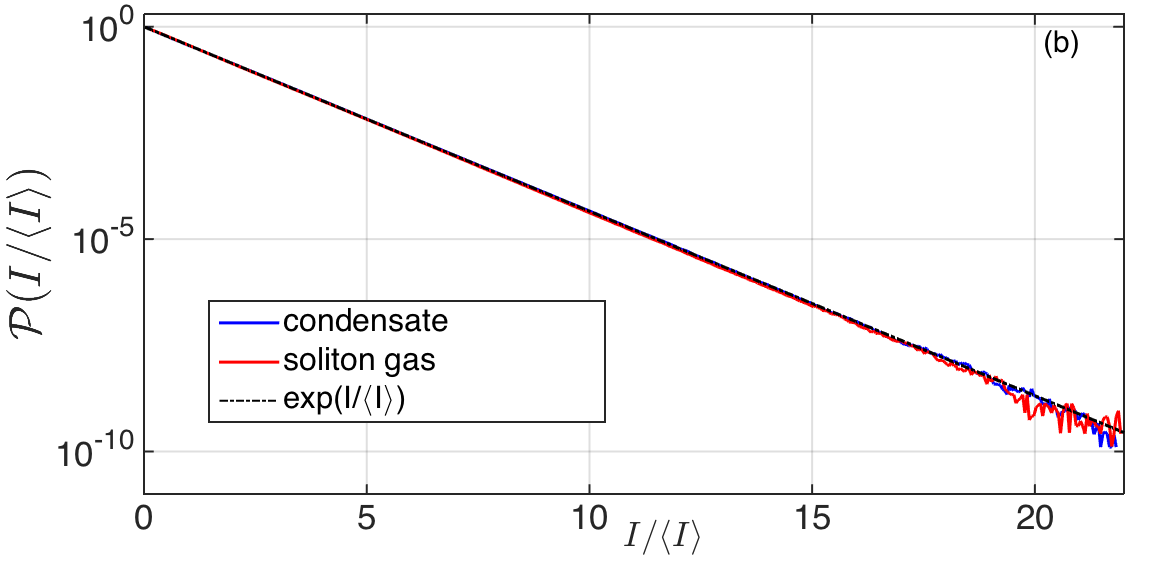}
\\
\includegraphics[width=0.99\linewidth]{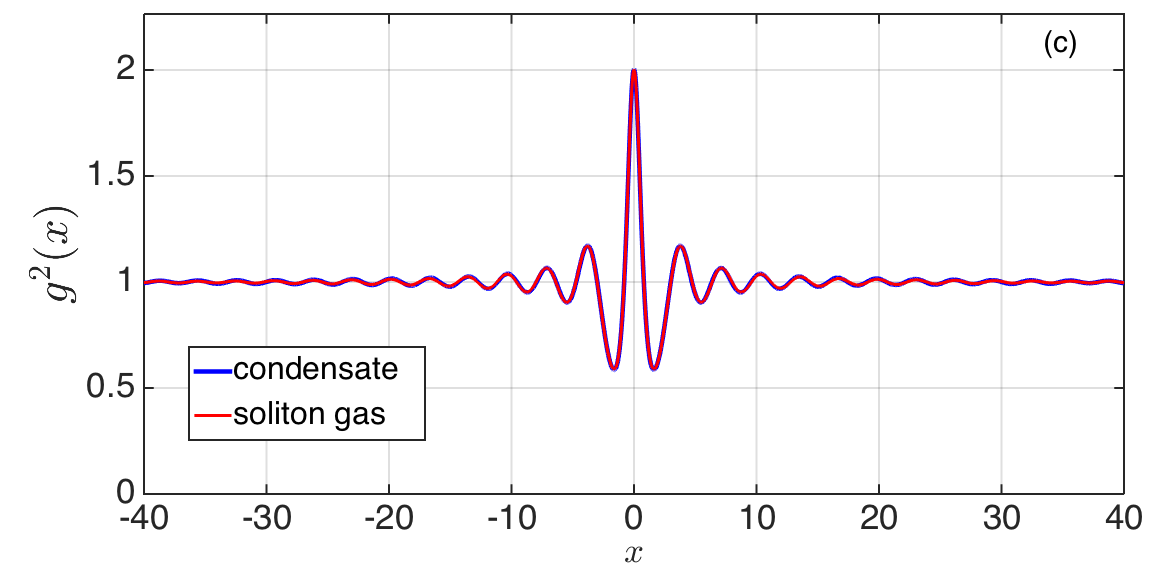}
 \caption{\label{Figure04}
Comparison of ensemble averaged statistical characteristics of the asymptotic state of the MI development and random phase $128$-SSs.
(a) Wave action spectrum $S_k$. (b) The PDF $\mathcal{P}(I)$. (c) Second order degree of coherence (autocorrelation function of intensity) $g^{(2)}(x)$}
\end{figure}

The wave-action spectrum, 
\begin{equation}\label{wave-action-spectrum}
S_{k} \propto \langle|\psi_{k}|^{2}\rangle,\quad \psi_{k} = \frac{1}{L}\int_{-L/2}^{L/2}\psi\, e^{-ikx}\,dx ,
\end{equation}
of the asymptotic state of the MI and of the considered $128$-SS soliton gas coincide with excellent accuracy, as demonstrated in Fig. 4a.~\ref{Figure04}. Note that $S_k$ is  renormalized porportionnaly to the spatial extension of the field.

Moreover, soliton gas and noise-induced MI exhibit nearly identical PDF $\mathcal{P}(I)$ of the field intensity $I=|\psi|^{2}$ (Fig.~\ref{Figure04}.b). The PDF of the $N$-SS reproduces quantitatively the exponential distribution discovered earlier as the asymptotic characteristic of the unstable condensate~\cite{Agafontsev:15, Kraych:19bis}.

It has been shown very recently in ~\cite{Kraych:19bis} that the long term evolution of the MI is typified by an oscillatory structure of the second order degree of coherence (autocorrelation of the intensity) $g^{(2)}(x)$ :
\begin{equation}
g^{(2)}(x)= \frac{\bigl\langle \int_{-L/2}^{L/2} I(y,t) I(y-x,t)dy \bigr\rangle }{ \bigl\langle \int_{-L/2}^{L/2} I(y,t) dy \bigr\rangle^2 } \,.
\label{eq:g2}
\end{equation}
As can be seen from the Figs.~\ref{Figure04}.c the $N$-SS  reproduce accurately this remarkable oscillatory shape.  Note that $H_l$, $H_{nl}$, the PDF and the $g^{(2)}$ functions of the $N$-SS were computed in the central part of the soliton gas. More precisely, we used the region $x \in [-150;150]$ where the ensemble-average wavefield intensity of the $N$-SS is uniform and very close to unity, that allows us to mitigate the edge effects.

\begin{figure}
\centering
\includegraphics[width=3.4in]{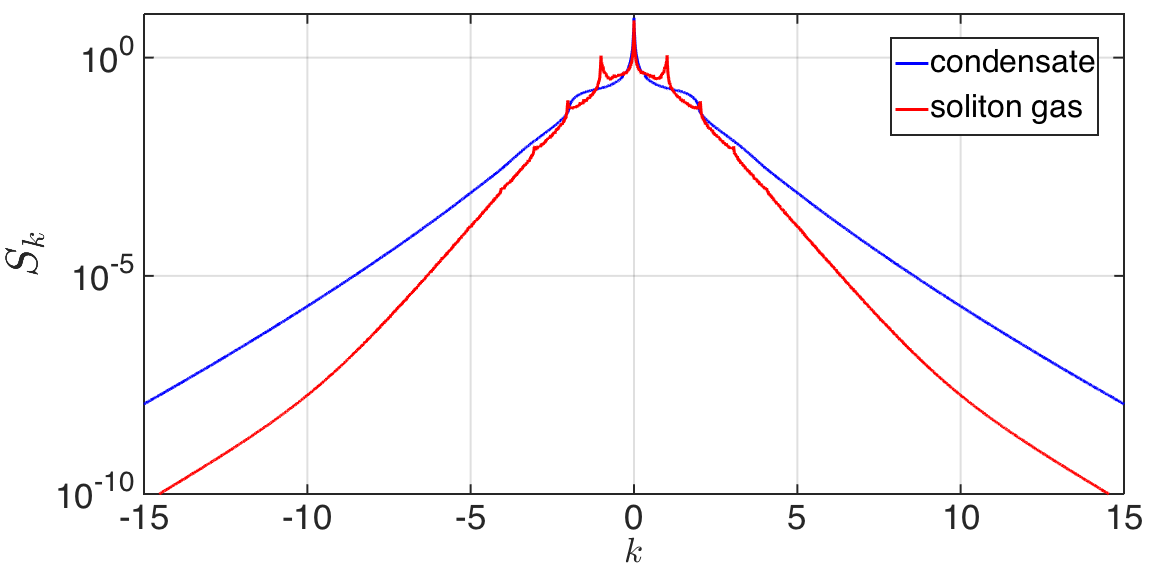}
\caption{\label{Figure05}
Example of wave action spectrum for soliton gas having equidistant distribution of $\lambda$.}
\end{figure}

As we have just shown, the asymptotic stage of the noise-induced MI is accurately modelled by a soliton gas with a special distribution of the IST eigenvalues $\lambda$, coinciding with the semi-classical distribution \eqref{eq:EIGdistr} for the discrete spectrum of the box potential~\cite{Zakharov:72}. This distribution  has been recently shown to describe the  density of states in the bound state soliton gas at critical density~\cite{El:19}.   Our numerical investigations reveal that spectral (Fourier) and statistical properties of soliton gas made of stochastic $N$-solitons bound states are very sensitive to the exact distribution of the IST eigenvalues.

 To illustrate the latter point, we compare in Fig.~\ref{Figure05} the typical (Fourier) spectrum of the asymptotic state of MI (identical to Fig.~\ref{Figure04}.a)  with the  spectrum of the  bound state soliton gas made of a $N$-SS with $N=128$, random phases and {\it equidistant} eigenvalues in the interval  $[\pi/4-a;\pi/4+a]$:
\begin{equation}\label{eq:EIGfixedSG3}
\lambda_n = i\, \beta_n = i\, \bigg[ \frac{\pi}{4} + 2 a \frac{n}{N} \bigg],  \ \ \text{where }a=0.025
\end{equation}

In sharp contrast to the  ``Weyl'' soliton gas studied above, the Fourier spectra of MI and of the soliton gas having the eigenvalue distribution~(\ref{eq:EIGfixedSG3}) strongly differ.
The detailed investigation of different soliton gases is beyond the scope of the paper; note however that the spectrum of soliton gas presented in Fig.~\ref{Figure05}
resembles the one of unstable cnoidal waves studied in ~\cite{Agafontsev:16}.
This example shows that the agreement between MI and a soliton gas (Fig.~\ref{Figure05}) is allowed by the careful choice of the density of states $f(\lambda)$. \\

In this paper we have demonstrated that the spectral and statistical properties of the asymptotic stage of MI precisely coincide with the ones of some specific soliton gas. This soliton gas can be constructed with exact $N$-soliton solutions of the 1D-NLSE having large values $N$ and one specific distributions of IST eigenvalues computed in the semi-classical limit~\cite{Zakharov:72}. As it could be expected, the long-term statistical state of MI correspond to a full stochastization of IST phases. While in this paper we concentrate on the asymptotic stage of the noise-induced MI, the proposed soliton gas framework may be useful in the understanding of the randomisation of the phases in the early stage of the MI.

We believe that our work  opens a new promising direction in the   theory of integrable turbulence  by establishing a link between the modulational instability and the soliton gas dynamics.  One of the most challenging problems is the rigorous derivation of the normal distribution  for the complex field $\psi(x,t)$ that typifies the nonlinear stage of noise-induced MI.  

Our model is based on the well-known multi-soliton solutions and can be generalised to a broad class of integrable turbulence problems  when the (random) wave field is strongly nonlinear, so that the impact of non-solitonic content can be neglected in the asymptotic state ($t\to\infty$).  In this case, the general recipe to study asymptotic state is to build $N$-soliton solutions with the distribution $f(\lambda)$  of eigenvalues characterizing the field and {\it random phases} of the norming constants. 

 There is one important remark to make. As is  known, the nonlinear stage of the  modulational instability induced by small {\it harmonic} perturbations is characterized by the generation of Akhmediev breathers ~\cite{Akhmediev:85,Akhmediev:86,Akhmediev:09, Grinevich:18}. The picture is more complicated if the initial perturbation is a random noise,  leading to integrable turbulence  with various breather structures appearing only locally \cite{Wabnitz:14, Toenger:15, Agafontsev:15, Narhi:16}.  The local appearance of breathers has also been demonstrated in the context of multi-soliton interactions \cite{Tikan:17, El:16:dambreak}. Our approach  can shed light on the possible connection between soliton gases and breather gases \cite{El:05:prl, El:19, Akhmediev:IntegrableTurbulence:16}.

 Note finally that the mechanism underlying the  MI induced by noise studied here is {\it a priori} different from the MI induced by localized perturbations ~\cite{Zakharov:13,Kibler:15, Biondini:16,Biondini:16:prlmi,Gelash:18, Conforti:18, Kraych:19}. The local perturbations are studied whithin the framework of IST with nonzero background conditions, and the corresponding dynamics can be strongly influenced by the soliton content {\it of the perturbation} ~\cite{Zakharov:13,Biondini:16:prlmi, Conforti:18, Gelash:18}. In contrast, whithin the proposed soliton model of a fully developed MI, we assume that the random perturbations of the condenstate (at t=0) only induce random phases of the special bound state soliton gas at $t \to \infty$.

As we demonstrate in this work, the IST formalism for the wave fields with {\it decaying} boundary conditions can be successfully applied to describe accurately the asymptotic stationary state of the MI computed numerically by using {\it periodic} boundary conditions~\cite{Agafontsev:15}. For an integrable PDE the periodic boundary problem is a subject of periodic IST technique also known as finite gap theory~\cite{Belokolos, Tracy:88}. An important mathematical problem for the future studies is to explain the link between spatially periodic and localised descriptions of the MI in terms of the IST theory.


\section*{Acknowledgments}
Simulations were performed at the Novosibirsk Supercomputer Center (NSU).
This work has been partially supported by the Agence Nationale de la Recherche through the LABEX CEMPI project (ANR-11-LABX-0007)
and by the Ministry of Higher Education and Research, Nord-Pas de Calais Regional Council and European Regional Development Fund (ERDF) through the Contrat de Projets Etat-R\'egion (CPER Photonics for Society P4S).
The work on the construction of the multi-soliton ensembles reported in the first half of the work was supported by the Russian Science Foundation (Grant No. 19-72-30028 to AG, DA and VZ).  The work of GE was partially supported by
EPSRC grant EP/R00515X/1. The authors thank A. Tikan and F. Copie for fruitful discussions
%

\end{document}